\documentclass[prstab]{revtex4}
\usepackage{graphicx}
\usepackage{amsmath}
\usepackage{latexsym}
\usepackage{amsfonts}
\usepackage{amssymb}

\newcommand{\sgn}{\mathop{\rm sgn}}

\begin{document}

\begin{flushright}
CBN 02-08 \\
\end{flushright}

\title{Equilibrium Distribution and Tune Shift of Beams in a Linear Collider}
\author{Bjoern S. Schmekel}
\author{Joseph T. Rogers}
\affiliation{Cornell University, Department of Physics, Ithaca, New York 14853, USA}
\begin{abstract}
The evolution of two colliding beams in a linear accelerator can be described by two
coupled Vlasov equations. In \cite{alex} the case without external focusing was
considered. In this paper we derive the equilibrium distribution and the tune shift 
in the presence of external focusing. Motion is considered only in the vertical 
direction and the beams are presumed to be one-dimensional.
\end{abstract}

\maketitle

\section{Beam Evolution}

The beam-beam force due to the first (second) beam on the second (first) one is given by
\begin{eqnarray}
\frac{d y^{\prime}_{1,2}}{ds} =
-\frac{4\pi Nr_{e}}{L_{x}\gamma}\int_{-\infty}^{\infty}
d\overline{y}\sgn(y-\overline{y})\int_{-\infty}^{\infty}d\overline
{y}^{\prime}\psi_{2,1}(\overline{y},\overline{y}^{\prime})
\label{beambeamforce}
\end{eqnarray}
where $N$ is the particle density in a bunch and $r_{e}$ the classical
radius of the electron. The distributions of the beams $\psi_{1}$ and
$\psi_{2}$ are normalized to unity, i.e.
\begin{eqnarray}
\int_{-\infty}^{\infty}d\overline{y}\int_{-\infty}^{\infty}%
d\overline{y}^{\prime}\psi_{1,2}(\overline{y},\overline{y}^{\prime})=1
\label{normalization}
\end{eqnarray}
and are assumed to be one-dimensional with horizontal width $L_{x}$. Motion is
considered only in the vertical direction. Then $\psi_{1,2}$ satisfy the Vlasov
equations
\begin{eqnarray}
\frac{\partial \psi_{1,2}}{\partial s} + y^{\prime} \frac{\partial \psi_{1,2}}{\partial y} -
\left ( K(s)y + \frac{4\pi Nr_{e}}{L_{x}\gamma} \int_{-\infty}^{\infty}d\overline{y}
\sgn(y-\overline{y})\int_{-\infty}^{\infty}d\overline
{y}^{\prime} \psi_{2,1}(\overline{y},\overline{y}^{\prime},s) \right )
\frac{\partial \psi_{1,2}}{\partial y^{\prime}} = 0
\label{vlasov}
\end{eqnarray}
\section{Equilibrium Distribution}
We are interested in finding an equilibrium (i.e. time-independent) distribution $\psi_0$. 
The ansatz 
\begin{eqnarray}
\psi_1(y,y^{\prime},s) = \psi_2(y,y^{\prime},s) = \psi_0(y,y^{\prime})
\end{eqnarray}
simplifies eqn. \ref{vlasov} significantly. 
\begin{eqnarray}
y^{\prime} \frac{\partial \psi_0}{\partial y} -
\left ( K(s)y + \frac{8\pi Nr_{e}}{L_{x}\gamma} \int_{-\infty}^{y}d\overline{y}
\int_{-\infty}^{\infty}d\overline
{y}^{\prime} \psi_0(\overline{y},\overline{y}^{\prime}) \right )
\frac{\partial \psi_0}{\partial y^{\prime}} = 0
\end{eqnarray}
This equation can be solved easily for a constant focusing function $K(s)=K$ :
\begin{eqnarray}
\psi_0(y,y^{\prime}) = f \left ( y^{\prime 2} + Ky^2 + \frac{16 \pi N r_e}{L_x \gamma}
\int dy \int_{-\infty}^{y} d \overline{y} \int_{-\infty}^{\infty} d\overline
{y}^{\prime} \psi_0(\overline{y},\overline{y}^{\prime}) \right )
\end{eqnarray}
where $f$ is any arbitrary function whose derivative exists. This equilibrium distribution $\psi_0$
must satisfy the normalization condition eqn. \ref{normalization}. We choose an exponential 
function because the beams going into the collision from a damping ring have an approximately
Gaussian distribution. Dividing both sides by the derivative with respect to 
$y^{\prime}$ and taking the derivative with respect to $y$ we obtain the following ordinary
differential equation 
\begin{eqnarray}
\frac{\partial^2 Y}{\partial y^2} Y - \left ( \frac{\partial Y}{\partial y} \right )^2 =
-2a Y^2 \left [ K + \sqrt{\frac{\pi}{a}} \xi Y \right ]
\label{equilibrium}
\end{eqnarray}
where $Y \equiv Y(y)$, $\xi \equiv \frac{8\pi N r_e}{L_x \gamma}$ and the ansatz 
\begin{eqnarray}
\psi_0(y,y^{\prime}) = Y(y)e^{-ay^{\prime 2}}
\label{yp}
\end{eqnarray}
has been used. The function $Y(y)$ satisfies the integral relation
\begin{eqnarray}
\int^{Y(y)} \frac{1}{\zeta \sqrt{-4 \xi \zeta \sqrt{a\pi} -4aK \ln(\zeta) + 
c_1}} d \zeta = y + c_2 
\end{eqnarray}
Depending on the constants $Y(y)$ describes single Gaussian-like functions,
a nearly periodic sum of Gaussian-like functions or oscillating functions.
Because Gaussian beams enter the collision, we are interested in the case
of a Gaussian-like $Y(y)$. Thus, we choose the ansatz
\begin{eqnarray}
\psi_0(y,y^{\prime}) = \frac{1}{2 \pi \sigma} e^{- \frac{J}{\sigma}}
\label{y}
\end{eqnarray}
where $J$ is defined for a constant focusing function $K=\frac{1}{\beta_0^2}$ as
\begin{eqnarray}
J=\frac{1}{2} y^{\prime 2} \beta + \frac{y^2}{2 \beta}
\end{eqnarray}
This leads to the identification $\sigma=\frac{\beta}{2a}$.
$\psi_0$ satisfies the normalization condition eqn. \ref{normalization}.
Substituting eqn. \ref{y} into eqn. \ref{equilibrium} yields a relation
for the tune shift
\begin{eqnarray}
\beta_0 = \frac{\beta^{3/2} (2 \pi \sigma )^{1/4} \sqrt{\sqrt{2 \pi \sigma} -
\xi \beta^{3/2} e^{- \frac{y^2}{2 \sigma \beta}}}}{\sqrt{2 \pi \beta \sigma} -
\xi \beta^2 e^{-\frac{y^2}{2 \sigma \beta}}} =
\beta \sqrt{\frac{\sqrt{2 \pi \sigma}}{\sqrt{2 \pi \sigma} - \xi \beta^{3/2} e^{-\frac{y^2}{2 \sigma \beta}}}}
\label{tuneshift}
\end{eqnarray}
If $\xi$ is sufficiently high the expression inside the root will become negative or - at a particular $\xi$ -
the denominator will vanish. Therefore, $\xi$ has to obey the inequality  
\begin{eqnarray}
\xi < \sqrt{\frac{2 \pi \sigma}{\beta^3}}
\end{eqnarray}
since the design tune has to be finite and real.

\section{Acknowledgments}
As always we would like to express our gratitude to Alex Chao from SLAC
for helpful discussions. This work was supported by the National Science
Foundation.

\end{document}